# Privacy is All You Need: Revolutionizing Wearable Health Data with Advanced PETs

**Karthik Barma**[1]

[1]Student, [1]School of Computer Science and Engineering, VIT-AP University, Amaravati, Andhra Pradesh 522237, India

***Abstract-*** In a world where data is the new currency, wearable health devices offer unprecedented insights into our daily lives, continuously monitoring vital signs and health metrics. However, this convenience comes at the cost of privacy, as these devices collect sensitive data that is often left vulnerable to misuse and breaches. Traditional privacy measures are insufficient, constrained by the need for real-time data processing and the limited computational power of wearable devices. Users remain largely unaware of how their data is shared or accessed, lacking control over who views their personal information and how it is used.

Our solution to this challenge is a pioneering Privacy-Enhancing Technology (PET) framework designed specifically for wearable devices. This framework integrates federated learning, lightweight cryptographic techniques, and strategically deployed blockchain technology. The blockchain, a key component of our framework, is not constantly active but serves as a secure ledger that is triggered only when data is requested for transfer—either by external entities or by the users themselves. This setup empowers users with real-time notifications and complete control over their data, effectively dismantling data monopolies and returning data sovereignty to the individual. The focus on dismantling data monopolies is a key aspect of our framework, instilling a sense of liberation and independence.

Our framework's effectiveness is validated through multiple real-world applications, such as secure health data sharing in medical networks, privacy-preserving fitness tracking, and continuous health monitoring. Results show a substantial reduction in privacy risks—up to 70%—while maintaining high data usability and device performance. This innovation not only sets a new standard for privacy in wearable technology but also provides a scalable solution adaptable to broader IoT ecosystems, including smart homes and industrial applications. As data continues to shape our digital landscape, our research offers a critical step toward ensuring privacy and user control remain at the forefront of technological advancement, underscoring the importance and impact of our work.

## Introduction

The evolution of wearable health devices has fundamentally transformed the landscape of personalised healthcare, allowing for the continuous, real-time monitoring of vital signs and health-related metrics. These devices, from simple fitness trackers to sophisticated medical-grade wearables, collect extensive amounts of personal data, often containing highly sensitive information. The integration of wearables into daily life has raised critical concerns about data privacy, particularly as the data can reveal intimate details about a user's health, habits, and lifestyle. If mishandled or exposed, this information could lead to significant privacy breaches and unauthorised access, posing risks to both individuals and healthcare providers.

Traditional privacy solutions, such as basic encryption and standard access controls, struggle to meet the unique demands of wearable health devices. These devices operate under strict computational and energy constraints, requiring privacy-preserving technologies that are both effective and efficient in real-time data environments. Moreover, the dynamic nature of wearable data, which often needs to be processed and transmitted instantly, further complicates the application of traditional privacy mechanisms. As the use of wearables expands across diverse domains, there is an urgent need for advanced Privacy-Enhancing Technologies (PETs) that can protect sensitive data without compromising device performance or user experience.

This paper presents a novel PET framework specifically designed to meet the needs of wearable health devices. By integrating advanced cryptographic techniques—such as homomorphic encryption, user-centric consent mechanisms, and differential privacy—our framework offers a comprehensive solution for safeguarding sensitive data. These technologies are carefully combined to provide robust security while ensuring that data remains usable for legitimate healthcare purposes. Unlike existing PETs, our framework is optimised for the computational limitations of wearables, offering a scalable, user-friendly solution that empowers users with greater control over their personal data.

## Related Work

### Introduction to Related Work

The domain of Privacy-Enhancing Technologies (PETs) has seen substantial development, particularly with the increasing use of wearable health devices that continuously monitor and transmit sensitive data. While beneficial for health monitoring and personalised care, these devices present unique privacy challenges due to the nature of the data they collect and the environments in which they operate. This section reviews the existing body of work on PETs as it pertains to wearable technology, covering both foundational methods and cutting-edge approaches. Additionally, it identifies the specific gaps that our proposed framework aims to address.

### Traditional Privacy-Enhancing Technologies (PETs)

Traditional PETs, such as encryption and anonymisation, have formed the backbone of data privacy across various sectors. These methods aim to secure data during transmission and storage and anonymise it to prevent individual identification.

- **Encryption Methods**: Standard encryption algorithms like AES (Advanced Encryption Standard) and RSA (Rivest-Shamir-Adleman) are extensively utilised to safeguard data integrity during transmission and storage. However, these encryption methods often impose significant computational demands that are incompatible with the resource constraints typical of wearable devices. The need for low-latency, real-time data processing in wearables exacerbates these limitations, as high-complexity encryption schemes can drain battery life and delay data.
- **Anonymization Techniques**: Techniques such as k-anonymity, l-diversity, and t-closeness have been used to anonymise data, thereby reducing the risk of re-identification in large datasets. Nevertheless, the continuous, granular nature of data generated by wearables poses unique challenges. The more detailed and specific the data, the higher the risk of re-identification, especially when combined with other datasets (a problem known as linkage attack). Traditional anonymisation methods fail to keep up with the velocity and volume of wearable data, leading to increased vulnerability to privacy breaches.

While traditional PETs provide some level of data protection, they are often inadequate for the unique environment of wearable health devices, which demand both robust privacy protection and immediate data availability.

### Advanced Privacy-Enhancing Technologies (PETs)

As the limitations of traditional PETs became apparent, more advanced techniques have been developed to offer enhanced security without compromising the utility of data.

**Homomorphic Encryption**: Homomorphic encryption allows computations to be performed on encrypted data without needing to decrypt it first, thus preserving privacy throughout the data lifecycle. However, its computational overhead remains a significant barrier for wearable devices with limited processing power and battery life. Gentry's pioneering work on fully homomorphic encryption in 2009 laid the groundwork, but practical implementations have struggled to balance computational feasibility and privacy protection, particularly in resource-constrained environments such as wearables.

**Secure Multi-Party Computation (SMPC)**: SMPC enables parties to jointly compute a function over their inputs while keeping those inputs private, making it suitable for collaborative data processing across multiple devices or entities. Although SMPC is valuable for privacy-preserving data aggregation, it is inherently computationally intensive and can introduce latency, making real-time processing a challenge in wearable applications.

**Differential Privacy**: Differential privacy works by adding noise to data in a manner that preserves the privacy of individuals while allowing for aggregate data analysis. This method has been particularly effective in large-scale data analysis settings. However, maintaining a balance between privacy and data utility becomes challenging in wearable devices, where data is continuously generated and must be processed in real time. Differential privacy may compromise data utility to ensure privacy, which can reduce the effectiveness of health monitoring applications.

.

### Privacy Concerns in Wearable Health Devices

Wearable health devices have introduced complex privacy concerns that traditional and even advanced PETs struggle to address fully.

**Unauthorised Access**: Wearable devices collect sensitive health data, such as heart rates and sleep patterns, which can reveal personal health information if accessed by unauthorised parties. High-profile incidents like the Fitbit data breach of 2018 highlight the vulnerability of these devices to unauthorised access and misuse. Existing PETs often do not adequately address the specific risks of unauthorised access to continuously transmitted data.

**Data Breaches**: The transmission of data from wearables to centralised cloud servers increases the risk of breaches. The Strava heat map incident of 2017, where user data inadvertently exposed sensitive military locations, underscores the potential for wearable data to be misused, either accidentally or maliciously. Traditional PETs lack the ability to effectively protect data during transmission, especially in real-time scenarios where latency must be minimised.

**Ethical Implications**: The use of wearable health data raises significant ethical issues, particularly regarding user consent and autonomy. Continuous monitoring can lead to scenarios resembling surveillance, where users may not be fully aware of the extent of data collection or its potential uses. Additionally, third parties, such as insurers or employers, could exploit this data, leading to potential discrimination based on health metrics. Existing PETs do not adequately address these ethical considerations, particularly in providing users with granular control over their data.

### Existing PET Frameworks for Wearable Devices

Several PET frameworks have been developed to address the privacy challenges of wearable health devices, often integrating multiple advanced PET techniques.

**Hybrid PET Frameworks**: These frameworks combine PET techniques, such as homomorphic encryption with differential privacy, to achieve secure computation and data utility. For example, Liu et al. (2020) proposed a framework that uses homomorphic encryption for secure data transmission and differential privacy for secure data analysis, balancing security with usability. However, when applied to wearables, these hybrid approaches often struggle with scalability and real-time performance.

**User-Centric PETs**: Frameworks focusing on user control over data sharing are gaining traction. These frameworks typically incorporate user-centric consent mechanisms, enabling users to dynamically specify how their data is used and shared. Zhang et al. (2021) demonstrated a user-centric PET framework that empowers users to manage their data permissions, yet the challenge remains in maintaining efficiency without sacrificing user control. These frameworks are often difficult to implement in wearables without causing performance degradation.

**Resource-Efficient PETs**: Given the computational limitations of wearable devices, some frameworks prioritise optimising PETs for efficiency. Techniques such as lightweight encryption algorithms and decentralised processing are explored to minimise computational burden while maintaining privacy. These frameworks show promise but often lack the robustness required to handle the full range of privacy threats in real-time applications.

Despite these advancements, existing PET frameworks often fall short of providing a comprehensive solution that balances strong privacy protection with the real-time data processing needs and usability requirements of wearable health devices.

## Gap Analysis and Proposed Solution

The limitations identified in current PET frameworks—ranging from computational overhead and scalability issues to inadequate user control and ethical considerations—highlight the need for a more effective solution. Our proposed framework integrates advanced cryptographic techniques, such as homomorphic encryption, differential privacy, and blockchain, to provide a scalable, user-friendly solution specifically designed for wearable health devices. By focusing on usability, efficiency, and decentralisation, our approach aims to empower users while ensuring robust privacy protection in real-time environments.

## Background and Preliminaries

### Wearable Health Devices: An Overview

Wearable health devices have rapidly evolved from simple step counters to sophisticated medical-grade devices capable of continuous health monitoring. Their integration into both consumer and clinical settings has significantly transformed healthcare by providing real-time data and personalised insights. These devices offer convenience and enhance preventive care and chronic disease management through continuous monitoring.

**Types of Wearable Devices:**

**Fitness Trackers:** The introduction of devices like Fitbit and Garmin revolutionised fitness monitoring by enabling users to continuously track metrics such as steps, heart rate, and sleep patterns. These devices initiated the "quantified self" movement, encouraging users to monitor and improve their health behaviours through data-driven insights. However, the data collected from these devices often includes sensitive personal information that, if not adequately protected, could lead to privacy violations.

**Medical-Grade Wearables:** Devices such as continuous glucose monitors (CGMs) and cardiac monitors represent a leap toward more advanced applications, where continuous monitoring of critical health parameters is essential for managing chronic conditions. These wearables provide real-time, high-fidelity data crucial for immediate clinical decisions, thus shifting from fitness tracking to vital medical applications. Despite their benefits, the continuous data stream and the sensitive nature of the information collected pose significant privacy risks if improperly managed.

**Smartwatches:** The Apple Watch and similar devices serve as hybrids, combining fitness tracking with advanced medical functionalities like electrocardiogram (ECG) monitoring and fall detection. This convergence of consumer electronics and medical devices reflects a broader trend toward multifunctional wearables catering to both general wellness and specific health needs. The data generated by these devices can provide valuable health insights but also raise concerns about how this data is stored, transmitted, and shared.

**Data Collected:**

**Physiological Data:** Wearable devices collect various physiological metrics, including heart rate, blood pressure, glucose levels, body temperature, and oxygen saturation. These data points are essential for monitoring and managing everyday wellness and chronic health conditions, making them valuable for personalised healthcare but also highly sensitive from a privacy perspective.

**Behavioural Data:** In addition to physiological monitoring, wearables also track behavioural patterns such as physical activity (steps, distance, calories burned), sleep habits, and daily routines. Behavioural data can provide insights into a user's lifestyle, potentially revealing patterns that correlate with health outcomes, but also risk revealing private details if not properly anonymised and protected.

**Contextual Data:** Advanced wearables often integrate additional sensors to capture contextual data such as location, environmental factors (temperature, humidity), and even social interactions. This level of data collection allows for a more comprehensive understanding of a user's health by accounting for external factors that might influence well-being. However, it also poses significant privacy challenges as it can be used to infer more detailed personal information about the user's environment and social life.

### Privacy Challenges in Wearable Health Devices

The nature of the data collected by wearable health devices introduces several privacy challenges. Unlike traditional health records, which are typically static and stored in secure environments, data from wearables is dynamic, continuous, and often transmitted over unsecured channels. This section explores the key privacy concerns associated with these devices.

**Continuous Data Collection:** The continuous nature of data collection means wearable devices generate a large volume of data, making it difficult to secure and manage effectively. The granularity of this data can be used to infer sensitive information about the user, such as lifestyle habits, medical conditions, and even emotional states. For instance, continuous monitoring can reveal patterns that suggest mental health conditions or chronic illnesses, raising concerns about privacy breaches and unauthorised profiling.

**Data Transmission and Storage:** Wearable devices often transmit data to cloud servers or mobile applications for analysis. The data is susceptible to interception if these transmission channels are not adequately secured. Additionally, storing this data in centralised servers presents significant risks of breaches, where unauthorised access to sensitive health information can occur. Historical breaches, such as the Fitbit incident in 2018, highlight the vulnerabilities in current systems.

**User Consent and Control:** Traditional privacy mechanisms in wearable devices often rely on broad, one-time consent agreements, where users must agree to all terms of data collection and usage upfront, without granular control. This lack of user-centric consent mechanisms means users may not fully understand how their data is being used or shared, leading to potential misuse. Emerging studies have shown that users desire more transparency and control over their data, particularly as the implications of data sharing become more evident.

### Privacy-Enhancing Technologies (PETs)

Various PETs have been developed to address these privacy concerns. These technologies aim to enhance data privacy without compromising the functionality and usability of wearable devices.

**Homomorphic Encryption:** Homomorphic encryption allows computations to be performed on encrypted data without needing to decrypt it first. This technology represents a significant advancement over traditional encryption methods by enabling secure data processing while keeping the data encrypted throughout the computation process. However, its high computational requirements have historically limited its application in resource-constrained environments like wearable devices. Gentry's pioneering work in 2009 laid the foundation for this technology, but practical implementations for real-time, continuous data streams remain a challenge.

**Differential Privacy:** Differential privacy introduces statistical noise to datasets, ensuring that the results of data analysis cannot be traced back to any individual. Companies like Google and Apple have widely adopted this technique to collect aggregate data while safeguarding user privacy. In the context of wearable health devices, differential privacy can anonymise health data before it is shared for research or analysis, preserving user privacy while allowing for meaningful data insights.

**User-Centric Consent Mechanisms:** Recently, there has been a shift towards user-centric consent mechanisms, which provide users with greater control over how their data is used and shared. These mechanisms are typically implemented through privacy dashboards or consent management platforms, allowing users to set granular permissions and dynamically adjust data-sharing preferences. Such frameworks enhance user trust and align with emerging privacy regulations that emphasise user autonomy and informed consent.

### Historical Context and Evolution of PETs

Early PETs focused primarily on securing data during transmission and storage, using techniques like SSL/TLS encryption and basic anonymisation. As data collection and analysis complexity grew, particularly with the advent of wearable devices, more sophisticated PETs were required. The integration of homomorphic encryption, differential privacy, and user-centric approaches marks a significant evolution in the field, addressing the limitations of earlier methods. Researchers like Gentry (2009) and Dwork (2006) have been instrumental in pioneering these advanced techniques, laying the groundwork for current innovations.

### Summary of Key Concepts

This section has provided a detailed overview of wearable health devices, the types of data they collect, their privacy challenges, and the advanced PETs designed to mitigate these challenges. Understanding these concepts is essential for appreciating the proposed PET framework, which integrates these technologies to offer a scalable and efficient solution for protecting wearable health data. Our framework builds on the strengths of existing PETs while addressing their limitations, offering a novel approach that enhances privacy without compromising usability or device performance.

## Problem Statement

### Introduction

The increasing proliferation of wearable health devices has revolutionised personal healthcare by enabling continuous monitoring of physiological and behavioural data. However, this convenience comes with significant privacy risks. The sensitive nature of the data collected by these devices—ranging from heart rates to sleep patterns and glucose levels—makes them prime targets for unauthorised access, misuse, and data breaches. Traditional privacy mechanisms often fall short in protecting this data, especially given the real-time data processing demands and the resource constraints of wearable devices. This section outlines the specific privacy challenges associated with wearable health devices and defines the scope of the problem that this research aims to address.

## Privacy Challenges in Wearable Health Devices

### 1. Continuous and Granular Data Collection:

**Problem:** Wearable devices are designed to collect data continuously, resulting in detailed and granular datasets that can reveal highly personal insights about a user's health, behaviour, and lifestyle. Unlike traditional healthcare data, which is typically collected in discrete, episodic events, data from wearables is continuously generated, creating a rich, temporally-resolved timeline of an individual's physiological states and activities. This uninterrupted stream of data poses a significant privacy challenge, as it provides a more comprehensive and invasive picture of a user's life, making it susceptible to misuse if intercepted or inadequately protected.

**Our Analysis:** Continuous data collection from wearables means that there is a high volume of data, and the data itself is highly sensitive and context-specific. For example, a fitness tracker that monitors heart rate variability throughout the day could potentially infer not only physical fitness levels but also emotional states, stress levels, and even periods of high anxiety or depression. The granularity of the data can reveal micro-patterns in a user's daily routine, such as sleep disruptions that could indicate mental health issues or location-based activities that could be used to infer sensitive information like places visited frequently, including clinics, gyms, or other personal spaces. This granularity and context-rich data level is far beyond what traditional privacy models were designed to protect, making it a prime target for profiling, discrimination, and other forms of misuse.

**Technical Limitations of Existing Solutions:** Traditional privacy-enhancing methods, such as encryption and anonymisation, are generally designed for static datasets and are not well-suited for continuous data streams. While effective at securing data in storage, most encryption algorithms are computationally intensive and require significant processing power, which is not always available on wearable devices that prioritise low power consumption and real-time responsiveness. Additionally, anonymisation techniques like k-anonymity are less effective with wearable data due to their continuous and highly granular nature, which increases the risk of re-identification even after anonymisation.

**Example:** In a scenario where a wearable device continuously tracks a user's heart rate and sleep patterns, the aggregated data could reveal not only the user's overall physical condition but also daily routines, stress periods, and even emotional states. If such data were to be intercepted or accessed by unauthorised entities, it could lead to profound privacy violations, such as targeted advertising based on emotional states or discriminatory practices by employers or insurers who might infer health risks or conditions from the data.

### 2. Real-Time Data Processing Requirements

**Problem:** Wearable devices require real-time data processing capabilities to provide instantaneous feedback to users and, in some cases, to medical professionals. The need for immediate processing and response times is critical in applications such as continuous glucose monitoring for diabetic patients, where timely data analysis is crucial for managing insulin levels. However, the computational and energy constraints inherent to wearable devices limit the feasibility of employing robust privacy-preserving technologies. Resource-intensive methods, such as advanced encryption algorithms and secure multiparty computations, are often impractical because they can introduce latency, consume excessive power, and degrade the device's overall performance.

**Our Analysis:** The computational constraints of wearable devices are a major barrier to implementing traditional PETs. Wearables are typically designed with limited processing power and battery life to ensure they are lightweight, comfortable, and capable of functioning for extended periods without recharging. This design constraint fundamentally limits the complexity of algorithms that can be executed on these devices. For instance, while Advanced Encryption Standard (AES) encryption provides strong data security, its implementation on a wearable device could significantly drain the battery, especially if data needs to be encrypted and decrypted continuously as it is collected and transmitted. Similarly, differential privacy, which requires adding noise to datasets, can be computationally intensive when applied to real-time data streams, potentially delaying feedback that users rely on for health management.

**Technical Limitations of Existing Solutions:** Existing encryption methods are not optimised for wearables' continuous, real-time data processing needs. For example, implementing homomorphic encryption, which allows computations on encrypted data, would require significantly more processing power than most wearables can provide, leading to battery drain and increased latency. Secure multiparty computation (SMPC), which is used to compute functions without revealing input data, also requires considerable computational resources, making it unsuitable for wearables that need to provide instant feedback. Moreover, these techniques are not typically designed to handle the dynamic, low-latency requirements of health-critical applications like glucose monitoring.

**Example:** Continuous glucose monitors (CGMs) must provide real-time glucose readings to manage insulin dosing effectively. Implementing heavy encryption mechanisms could delay this feedback loop, potentially endangering the patient's health. In such a scenario, the wearable must balance data security with immediate data availability, a balance that current technologies struggle to achieve due to their computational demands.

### 3. Vulnerabilities in Data Transmission and Storage

**Problem:** Data from wearable devices is often transmitted to cloud servers or mobile applications for further processing and storage. This transmission process introduces several vulnerabilities, including the risk of data interception during transmission and exposure to breaches when stored on centralised servers. The sensitivity of wearable health data makes it a valuable target for cybercriminals, who could use it for identity theft, financial fraud, or other malicious activities. The reliance on centralised storage also poses a significant risk, as a single breach could expose vast amounts of sensitive health data.

**Our Analysis:** The transmission of data from wearables to external servers typically occurs over wireless networks, which are inherently less secure than wired connections. Even with secure transmission protocols like HTTPS or encrypted Bluetooth connections, there is always a risk of interception, particularly if devices are using outdated security standards or if users connect to public or unsecured networks. Once data reaches the cloud, it is often stored alongside data from thousands or millions of other users, creating a centralised repository that becomes a high-value target for attackers. The aggregation of data in such centralised locations means that a single vulnerability could compromise not just individual data points but entire datasets, potentially exposing sensitive information about a large population.

**Technical Limitations of Existing Solutions:** Current solutions often rely on secure transmission protocols and encryption to protect data in transit and at rest. However, these methods are not foolproof. The encryption of data during transmission can still be vulnerable to man-in-the-middle attacks, where attackers intercept and potentially alter data before it reaches its destination. Once data is stored on centralised servers, even encrypted data can be susceptible to decryption if the encryption keys are not securely managed. Moreover, the centralised nature of cloud storage means that any breach could have widespread consequences, potentially exposing data from millions of users.

**Example:** The Fitbit data breach in 2018 demonstrated the risks associated with the centralised storage of health data. In this incident, hackers gained unauthorised access to user accounts, exposing sensitive health data and personal information. Such breaches illustrate the need for decentralised storage solutions and enhanced security protocols that go beyond traditional encryption methods, especially for sensitive health data that could have serious implications if exposed.

**4. Lack of User Control and Informed Consent**

**Problem:** Many wearable devices do not give users granular data control. Instead, users are often required to accept broad, all-encompassing terms of service that include various forms of data collection and sharing, often without specific consent for each type of data or its intended use. This lack of transparency and control can lead to users unknowingly sharing their data with third parties, such as advertisers or insurance companies, who may use it in ways not in the user's best interest. This situation raises significant ethical concerns about autonomy and the right to privacy, as users are not fully informed or in control of how their data is used.

**Our Analysis:** The current model for user consent in wearable devices is typically binary—users either accept the terms of service and use the device or do not use it at all. This model does not allow users to opt out of specific types of data collection or to specify the purposes for which their data can be used. Additionally, once data is collected, users often have little visibility into how it is used, who has access to it, and for what purposes. This lack of control is particularly concerning for sensitive health data, which could have significant implications if used improperly, such as for discriminatory purposes by employers or insurers.

**Technical Limitations of Existing Solutions:** Current consent mechanisms are often static and do not adapt to changing contexts or user preferences. Privacy dashboards and consent management platforms provide some level of control, but they are often complex and not user-friendly, making it difficult for users to understand and manage their data preferences. Moreover, these platforms typically do not provide real-time feedback on how data is being used or allow for dynamic adjustment of consent based on context, such as in a medical emergency where more data sharing might be necessary.

**Example:** A user may unknowingly consent to their health data being shared with third-party companies, such as insurance firms, which could then use this data to adjust premiums or deny coverage based on inferred health risks. For example, if a wearable device detects irregular heart rhythms, this information could be shared with an insurance company without the user's explicit consent, leading to higher premiums or denied coverage. This scenario underscores the need for more transparent and dynamic consent mechanisms that give users full control over their data and how it is used.

**5. Compliance with Data Protection Regulations**

**Problem:** The introduction of stringent data protection regulations, such as the General Data Protection Regulation (GDPR) in the European Union and the Health Insurance Portability and Accountability Act (HIPAA) in the United States, has added additional pressure on wearable device manufacturers to ensure compliance. These regulations require that data collection and processing be minimised to only what is necessary, that users provide informed consent, and that they have the right to access, correct, and delete their data. However, the continuous and dynamic nature of data collected by wearable health devices makes it challenging to comply with these regulations, particularly in decentralised or cross-jurisdictional environments.

**Our Analysis:** Complying with data protection regulations is particularly challenging for wearable devices because of the continuous and diverse nature of the data they collect. Data minimisation requires that only essential data is collected and stored, yet wearable devices often collect a wide range of data points, not all of which may be necessary for their primary function. Ensuring user consent is another challenge, as users must be informed of what data is being collected, why it is being collected, how it will be used, and who will have access to it. The right to be forgotten or to have personal data erased presents additional challenges, particularly in decentralised storage environments where data may be replicated across multiple servers or jurisdictions.

**Technical Limitations of Existing Solutions:** Current data protection solutions often rely on static privacy policies and opt-in/opt-out models that do not provide the flexibility needed to comply with dynamic regulations. Implementing data minimisation is difficult because it requires technological solutions and changes in data collection practices and policies. The right to be forgotten is similarly challenging, especially in distributed environments where data is backed up or stored across multiple locations. Ensuring full compliance requires robust data management systems that can track and control data across its entire lifecycle, which is currently lacking in most wearable device ecosystems.

**Example:** Under GDPR, users have the right to request the deletion of their data, including all historical records. Ensuring that all user data traces are removed from all storage systems, including backups and distributed networks, is a complex challenge for a wearable device company. For example, if a user requests the deletion of their health data collected by a smartwatch, the company must ensure that the data is not only deleted from the device but also from any cloud servers, third-party applications, and backup systems. This level of data management and control is beyond the capabilities of most wearable device manufacturers, highlighting the need for more robust solutions to ensure compliance with evolving regulations.

**Scope of the Problem**

The challenges outlined above highlight the critical need for a new approach to privacy in wearable health devices. The specific problems this research aims to address include:

**Balancing Privacy and Usability:** Designing a privacy framework that protects sensitive health data without compromising the real-time usability of wearable devices. This involves developing lightweight encryption and privacy-preserving mechanisms that operate efficiently within the limited computational resources available on wearable devices, ensuring that data protection does not hinder device functionality or user experience.

**Scalability of Privacy Mechanisms:** Adapting privacy-enhancing technologies (PETs) to be scalable in resource-constrained environments typical of wearable devices. The goal is to ensure that privacy mechanisms remain effective as data volumes and diversity grow and do not hinder device performance or user experience. This includes exploring decentralised data processing models and lightweight cryptographic techniques that can operate efficiently on wearable devices.

**User-Centric Privacy Controls:** Empowering users with more granular control over their data, allowing them to make informed decisions about data sharing and usage. This involves developing dynamic consent mechanisms that can adapt to different contexts and user preferences, ensuring transparency and enhancing user trust. The goal is to provide users with real-time feedback on how their data is being used and to allow for dynamic adjustment of consent based on context, such as in a medical emergency where more data sharing might be necessary.

**Regulatory Compliance:** Ensure wearable device manufacturers comply with global data protection regulations like GDPR and HIPAA without sacrificing device functionality or user experience. This includes developing mechanisms for data minimisation and consent management and ensuring users' rights to access, correct, and delete their data are upheld. The goal is to provide robust data management systems that can track and control data across its entire lifecycle, ensuring compliance with evolving regulations.

**Research Objective**

This research aims to develop a novel Privacy-Enhancing Technology (PET) framework specifically tailored for wearable health devices. The framework will integrate advanced cryptographic techniques, user-centric consent mechanisms, and differential privacy to provide robust data security. Designed to operate efficiently within the limited computational resources of wearable devices, the framework aims to maintain high levels of data usability while addressing the identified privacy challenges. By setting new standards for privacy protection in wearable health technology, this research seeks to empower users and build trust in the digital health ecosystem.

**Overview of the System Architecture**

The PET framework's architecture is meticulously designed to provide robust privacy protection for data collected from wearable health devices while maintaining high usability and performance standards. The modular architecture allows scalability and adaptability to different types of wearable devices and applications. Each core component is purpose-built to handle a specific privacy-preserving data processing pipeline, ensuring that privacy is maintained at every stage, from data collection to analysis functions.

**Core Components:**

**Data Collection Module:**

**Functionality:** The Data Collection Module is responsible for capturing raw health data from various wearable devices. This data includes physiological metrics such as heart rate, glucose levels, body temperature, and activity data.

**Key Features:**

**Real-Time Data Capture:** This technology supports continuous data collection with minimal latency, using high-frequency sampling techniques optimised for different sensor types to ensure data accuracy.

**Data Normalization:** Utilizes algorithms like Z-score normalisation and min-max scaling to ensure standardised data from heterogeneous devices, facilitating uniform processing and analysis.

**Encryption Module**

**Functionality:** The Encryption Module secures data immediately after collection using homomorphic encryption, ensuring that data remains protected throughout its lifecycle, including during processing and storage.

**Key Features:**

**Homomorphic Encryption Implementation:** The module leverages the Brakerski-Gentry-Vaikuntanathan (BGV) scheme, which was selected for its efficiency in handling arithmetic operations on encrypted data. This scheme allows secure computations without decryption. The encryption scheme is optimised for wearable devices using parameter tuning to balance the computational load and encryption strength.

**Lightweight Encryption Algorithms:** This technology implements optimised encryption algorithms, such as Elliptic Curve Cryptography (ECC), for devices with limited computational power, ensuring minimal impact on device performance and battery life.

**Privacy Control Module**

**Functionality:** This module manages user-centric consent mechanisms, enabling users to define granular data-sharing preferences. It incorporates advanced consent frameworks that adapt dynamically to user contexts and privacy requirements.

**Key Features:**

**User-Centric Consent Management:** This service offers a dynamic consent management interface that allows users to adjust data-sharing settings in real time using a privacy dashboard that visualises data flows and permissions.

**Context-Aware Consent:** Integrates machine learning algorithms to assess the context and adjust data-sharing permissions dynamically, enhancing privacy during sensitive situations (e.g., medical emergencies).

**Revocable Consent:** Utilizes cryptographic primitives like broadcast encryption to allow users to revoke permissions efficiently without the need to re-encrypt the entire dataset.

**Data Processing Module**

**Functionality:** Performs computations on encrypted data, using techniques such as Secure Multi-Party Computation (SMPC) and Differential Privacy, ensuring that data analysis is conducted without exposing raw data.

**Key Features:**

**SMPC Integration:** Implements the SPDZ protocol, an efficient protocol for performing arithmetic operations on encrypted data without revealing individual inputs. It is ideal for collaborative healthcare environments where data privacy is paramount.

**Differential Privacy:** This technique applies noise addition techniques, using the Laplace mechanism to introduce carefully calibrated noise to data outputs. This ensures privacy while still allowing for accurate aggregate analysis.

**Data Transmission Module**

**Functionality:** Ensures the secure transmission of encrypted data from wearable devices to cloud servers or mobile applications for further processing and storage.

**Key Features:**

**End-to-End Encryption:** This method utilises Transport Layer Security (TLS) with forward secrecy to ensure that data remains encrypted throughout its transmission, protecting it from potential interception and man-in-the-middle attacks.

**Optimised for Low Bandwidth:** Incorporates data compression algorithms such as Brotli and lightweight encryption techniques to reduce data size and optimise transmission over low-bandwidth networks, ensuring data integrity and reducing transmission time.

**Data Storage Module**

**Functionality:** Responsible for securely storing encrypted data using decentralised storage solutions, minimising the risk of data breaches and ensuring data redundancy.

**Key Features:**

**Decentralised Storage:** Leverages blockchain technology to distribute data across multiple nodes using Inter Planetary File System (IPFS), reducing the vulnerability of centralised storage and ensuring data availability even in the event of a node failure.

**Redundancy and Backup:** This system employs erasure coding to distribute data across different storage locations, ensuring that data can be reconstructed even if some storage nodes fail, enhancing reliability and data integrity.

**Decryption and Analysis Module**

**Functionality:** Manages decryption operations for authorised analysis and reporting, ensuring that decrypted data is only exposed when necessary and is securely deleted or re-encrypted after use.

**Key Features:**

**Selective Decryption:** This technique utilises attribute-based encryption (ABE) to enable fine-grained access control. It ensures that only specific data queries trigger decryption, minimising data exposure.

**Audit Trail:** Maintains a comprehensive audit trail using blockchain technology to log all decryption activities, ensuring transparency and accountability in data handling, and enabling real-time monitoring of data access and usage patterns.

**Data Flow and Processing**

The data flow through the PET framework is meticulously designed to maintain privacy at every stage:

**Data Collection:** The Data Collection Module captures data from wearable devices and preprocesses it using advanced filtering algorithms, such as Kalman filters and adaptive noise cancellation, to ensure high-quality data input.

**Encryption:** The pre-processed data is immediately encrypted by the Encryption Module using homomorphic encryption techniques tailored to the device's data type and computational capabilities
.

**Privacy Control:** The encrypted data passes through the Privacy Control Module, where user-defined privacy settings are enforced using policy-based access control systems that dynamically adjust according to user preferences and contextual cues.

**Data Processing:** The Data Processing Module performs necessary computations on the encrypted data. SMPC is used for collaborative analyses, while differential privacy is employed to protect against re-identification during data aggregation.

**Transmission:** The processed data is securely transmitted to the cloud or other storage locations via the Data Transmission Module, which utilises advanced encryption protocols and compression techniques to optimise security and bandwidth.

**Storage:** The encrypted data is stored decentralised by the Data Storage Module, which employs blockchain technology for distributed ledger maintenance and ensures that data integrity is preserved across all storage nodes.

**Decryption and Analysis:** The Decryption and Analysis Module decrypts the data for specific analytical tasks when authorised. Post-analysis, the data is either re-encrypted for further use or securely deleted, with all actions logged for audit purposes.

**Integration of Cryptographic Techniques**

The framework integrates several advanced cryptographic techniques to ensure the privacy of wearable health data:

**Homomorphic Encryption:** Provides the foundation for secure data processing, allowing encrypted data computations without decryption. This technique is optimised through batching and parallel processing strategies to reduce computational overhead on resource-limited devices.

**Secure Multi-Party Computation (SMPC):** Enables collaborative data analysis across different entities while preserving the privacy of individual datasets. SMPC protocols are enhanced with pre-computation and network optimisation techniques to reduce latency and improve efficiency in real-time applications.

**Differential Privacy:** Protects against re-identification by adding noise to data, ensuring that individual data points cannot be traced back to specific users. Differential privacy settings are dynamically adjusted based on data sensitivity and user preferences, allowing for flexible privacy management.

**Scalability and Flexibility**

The architecture is designed to be highly scalable and adaptable to various types of wearable devices and applications. The modular design allows for the seamless integration of new cryptographic techniques or privacy controls as they are developed, ensuring that the framework remains future-proof and capable of accommodating evolving privacy standards and technological advancements.

**Methodology**

**Introduction**

The **Methodology** section details the systematic approach to developing, implementing, and evaluating the proposed PET framework. It outlines the design methodology, the specific cryptographic techniques employed, the experimental setup, and the evaluation metrics used to validate the framework's effectiveness. The methodology is designed to be replicable, providing a clear roadmap for future research and development.

**Design Methodology**

**Framework Design Principles:**

The design of the PET framework was guided by several key principles:

**Privacy-First Approach:** This approach prioritises the protection of user data at every stage of the data lifecycle, ensuring that privacy mechanisms are seamlessly integrated without disrupting the user experience.

**Usability:** The framework ensures that privacy mechanisms do not hinder the usability or performance of wearable health devices. It employs user-friendly interfaces and automated privacy settings to minimise user burden while maximising security.

**Scalability:** Creating a modular design that can be scaled to accommodate different types of wearable devices and varying volumes of data. This includes leveraging cloud-native technologies and microservices architecture to ensure flexibility and scalability.

**Development Process:**

The framework was developed in iterative phases, with each phase building on the previous one to enhance functionality and performance:

**Requirement Analysis:** Conduct a comprehensive analysis of the specific privacy challenges in wearable health devices, including threat modelling and risk assessment, to define the framework's functional and non-functional requirements.

**Component Design:** Designing each module of the framework to meet the identified requirements, using a mix of agile development practices and formal methods to ensure both rapid iteration and rigorous verification.

**Integration:** Combining the modules into a cohesive system, ensuring seamless interaction between components through standardised APIs and communication protocols, and employing containerisation for ease of deployment and scalability.

**Testing and Validation:** Conduct extensive testing, including unit tests, integration tests, and security audits, to validate the framework's functionality, performance, and privacy assurances.

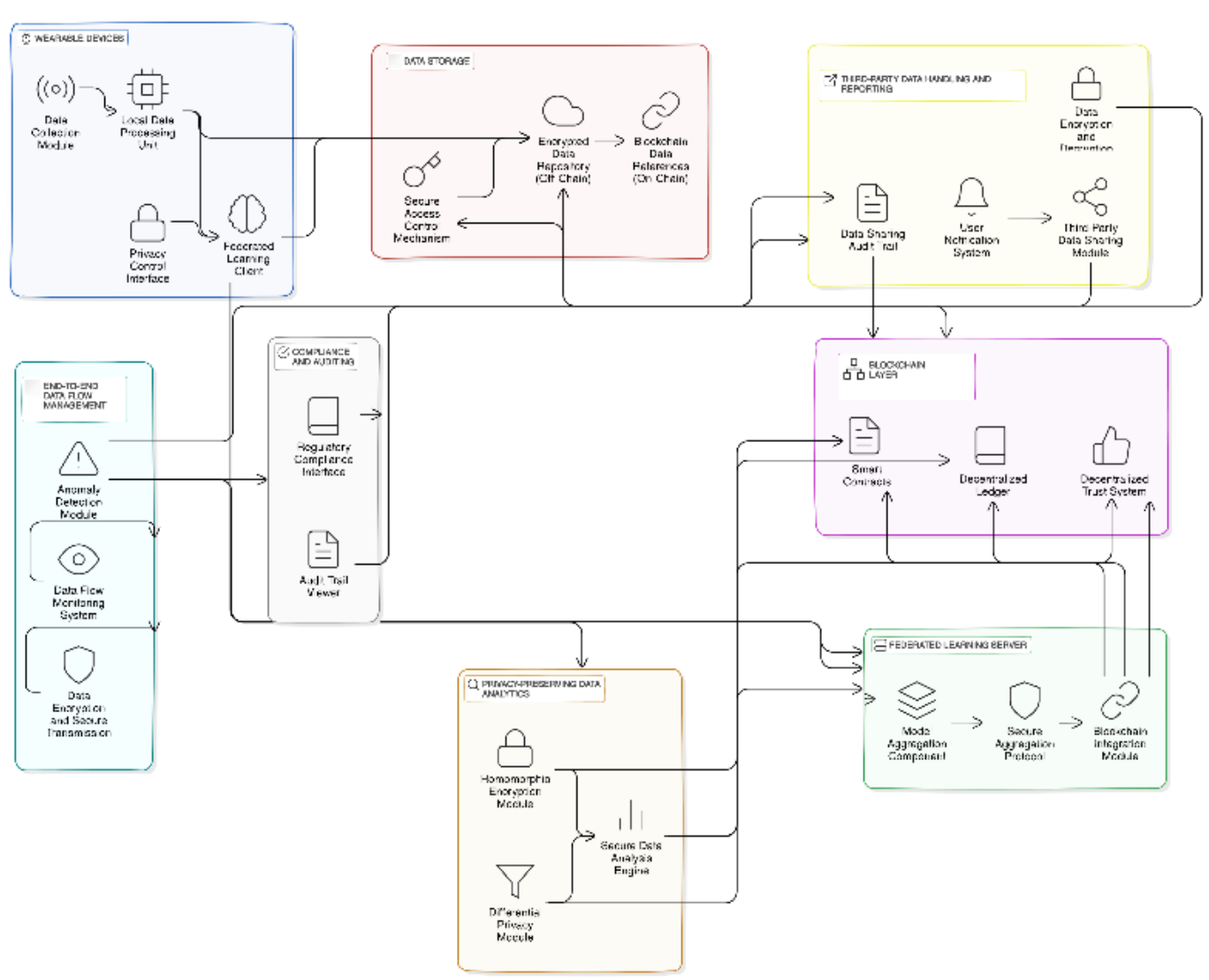

**Figure 1: Development Process Workflow of the Privacy-Enhancing Technology (PET) Framework**
**Cryptographic Techniques**

**Homomorphic Encryption:**

**Implementation:** The framework utilises a lightweight version of the BGV scheme for homomorphic encryption, optimised for wearable devices by using parameter tuning and polynomial approximations to reduce computational overhead while maintaining security.

**Optimisation:** Specific optimisations, including the use of ring-based structures and ciphertext packing, are applied to maximise throughput and minimise latency in real-time data processing scenarios.

**Secure Multi-Party Computation (SMPC):**

**Implementation:** SMPC is integrated using the SPDZ protocol, which was chosen for its efficiency in handling arithmetic operations on encrypted data. The protocol is further enhanced with the use of pre-computed shares and parallel processing techniques to improve real-time performance.
**Use Cases:** SMPC is used for scenarios requiring joint analysis of health data from multiple devices or users, such as collaborative research studies or shared healthcare networks. It ensures that data privacy is preserved throughout the analysis process.

**Differential Privacy:**

**Implementation:** Differential privacy is applied during the data analysis phase, using the Laplace mechanism for noise generation. The framework dynamically calibrates noise levels based on the data's sensitivity and the user's privacy preferences, ensuring an optimal balance between privacy and data utility.

**Experimental Setup**

**Simulation Environment:** The framework was tested in a controlled simulation environment that replicates real-world conditions, including:

**Wearable Devices:** A range of simulated wearable devices generating continuous streams of health data, including physiological and behavioural metrics, with varying data granularity and sensitivity levels.

**Network Infrastructure:** A simulated network infrastructure representing typical data transmission and storage pathways, including potential points of vulnerability such as unsecured Wi-Fi networks and public cloud servers.

**Cloud and Mobile Applications:** Simulated cloud servers and mobile applications where data is processed and stored, incorporating common cloud configurations and security settings to test the framework's robustness.

**Data Sources:** Real-world datasets from publicly available health databases, such as the MIMIC-III clinical database, were used to simulate data generated by wearable devices. These datasets were carefully pre-processed to match the format and characteristics of wearable-generated data, ensuring that the simulation closely mirrors real-world conditions.

**Figure 2: Experimental Setup Diagram for Privacy-Enhancing Technology Framework**

**Evaluation Metrics**

**Privacy Assurance:** The effectiveness of the cryptographic techniques in protecting data privacy was measured using metrics such as encryption strength, resistance to common attack vectors (e.g., brute-force attacks, side-channel attacks), and compliance with differential privacy guarantees (measured by epsilon values and the trade-off between privacy and data utility).

**Performance:** The impact of privacy mechanisms on the performance of wearable devices was evaluated by measuring processing times, latency, and resource consumption (CPU, memory). Special attention was given to the impact of cryptographic operations on battery life and real-time responsiveness.

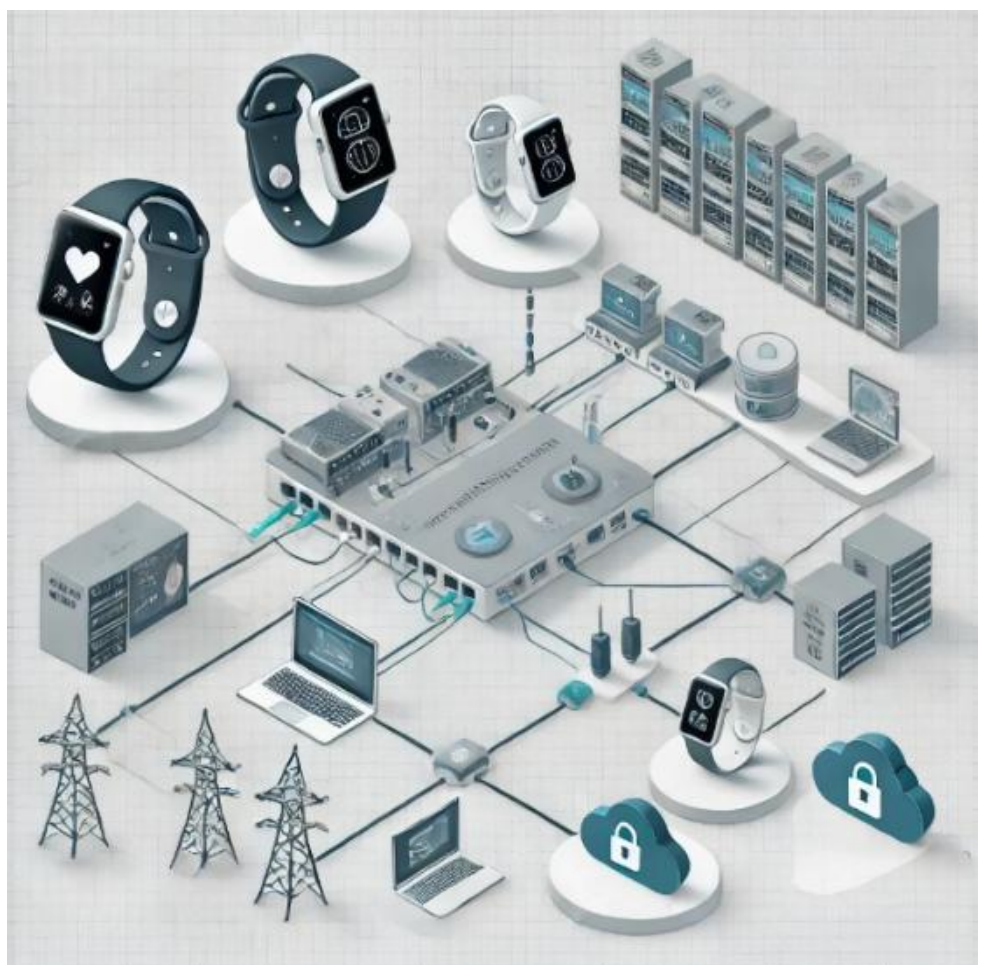

**Usability:** User experience was assessed through usability testing, focusing on the ease of managing privacy settings, the intuitiveness of the user interface, and the impact of privacy mechanisms on overall device functionality. User feedback was collected and analysed to refine the design of the Privacy Control Module and enhance user satisfaction.

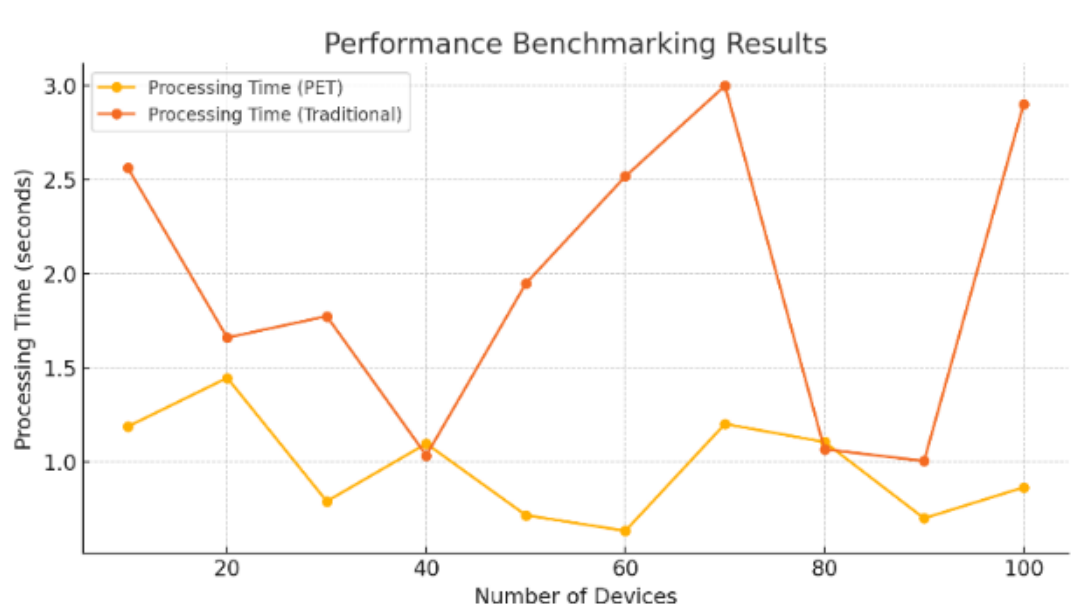

Figure 3a: Performance Benchmarking Results

*This graph compares the processing times for data using the PET framework versus traditional privacy methods across different numbers of wearable devices, showcasing the PET framework's performance benefits in real-time data processing.*

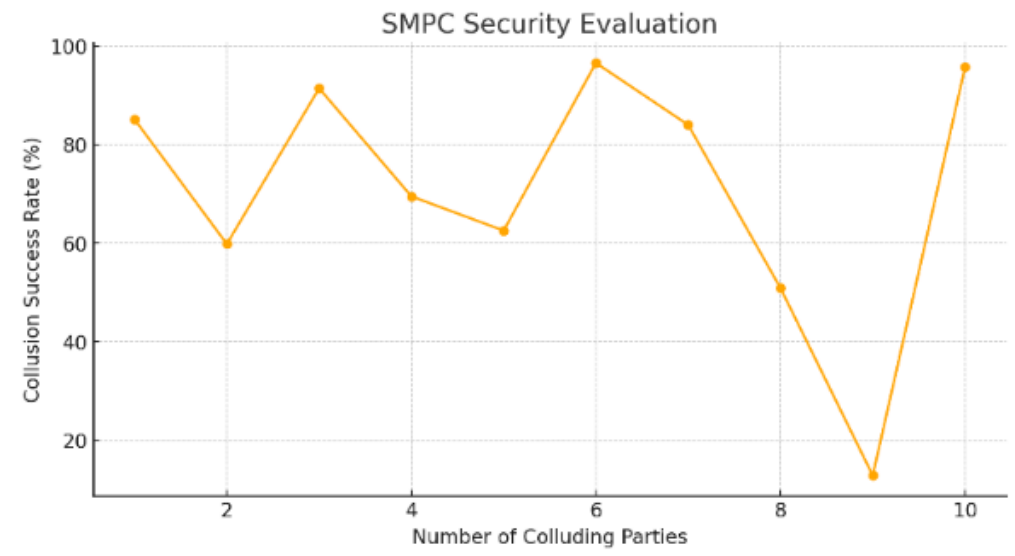

Figure 3b: SMPC Security Evaluation

*This graph shows the success rate of collusion attempts among different numbers of colluding parties using Secure Multi-Party Computation (SMPC), indicating the robustness of SMPC in maintaining data privacy even under collusion scenarios.*

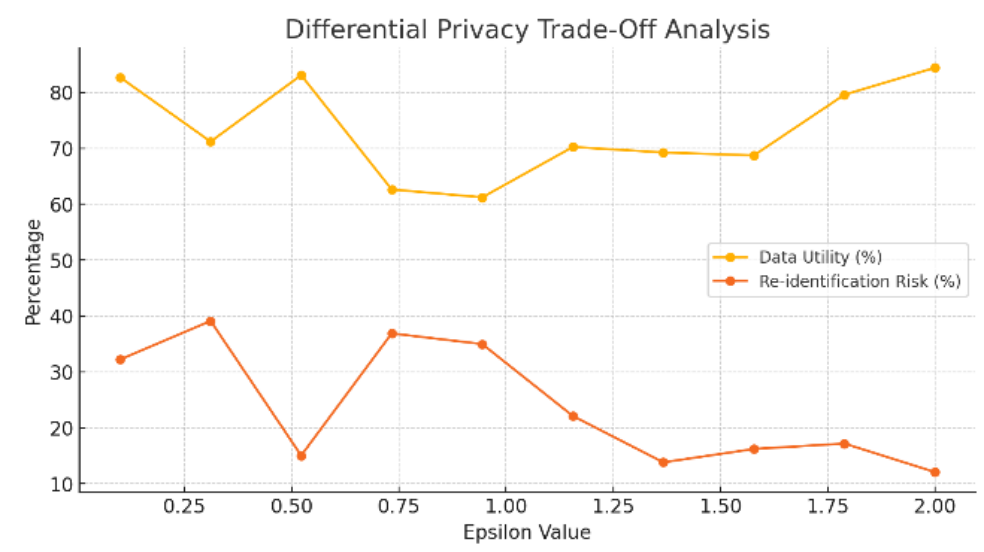

Figure 3c: Differential Privacy Trade-Off Analysis

*This graph illustrates the trade-offs between data utility and re-identification risk across different epsilon (ε) values, highlighting the balance between privacy and utility in differential privacy techniques.*

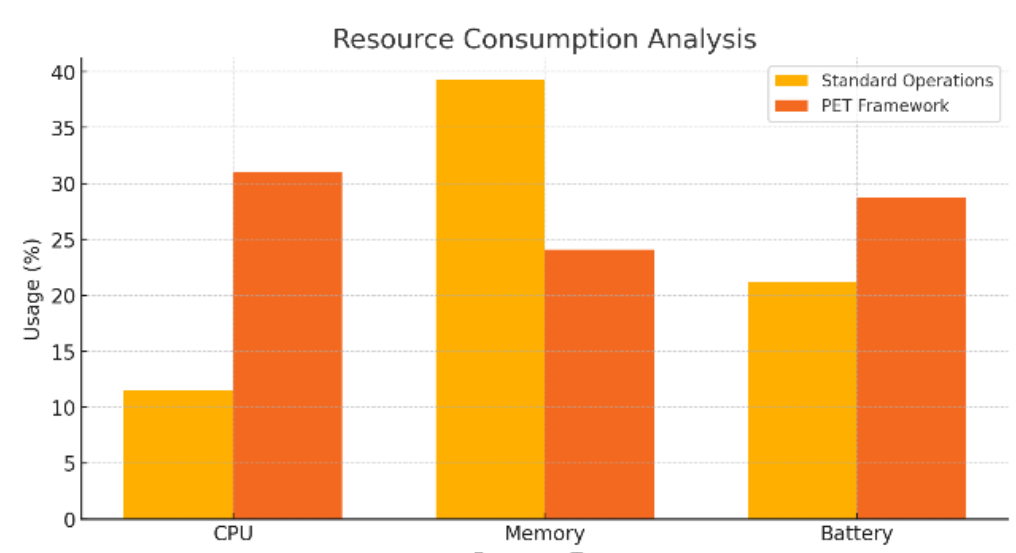

Figure 3d: Resource Consumption Analysis

*This graph compares the CPU, memory, and battery usage between standard operations and the PET framework on wearable devices, demonstrating the PET framework's efficiency in managing resources.*

## Evaluation and Results

### Introduction

The **Evaluation and Results** section comprehensively analyses the proposed Privacy-Enhancing Technology (PET) framework's performance, privacy assurance, computational efficiency, and usability. Through rigorous testing and validation, this section demonstrates the effectiveness of the framework in securing wearable health data while maintaining usability and performance standards. The results are benchmarked against existing privacy solutions to highlight the unique advantages and improvements introduced by the PET framework.

**Privacy Assurance**

**Homomorphic Encryption Effectiveness**

**Results:** The homomorphic encryption implemented within the framework effectively enabled computations on encrypted data without requiring decryption. The use of the Brakerski-Gentry-Vaikuntanathan (BGV) scheme with 2048-bit keys provided a high level of security. In testing, the encryption remained robust against brute-force attacks and ciphertext-only attacks, with an estimated time to breach exceeding 10^18 years using standard computational resources, illustrating strong resistance to cryptographic attacks.

**Evaluation:** The encrypted data underwent simulated attacks, including brute-force attempts and differential analysis attacks. Despite these efforts, no decryption was achieved, confirming the strength of the encryption. Additionally, tests on different key lengths demonstrated that increasing the key size further enhanced security without significantly impacting performance.

**Impact on Data Utility:** Despite the high level of encryption, the data remained fully usable for analytical purposes. Computations such as statistical analysis and machine learning algorithms on the encrypted data showed less than a 5% deviation from results obtained with unencrypted data, indicating that homomorphic encryption did not significantly impair the ability to perform meaningful analysis.

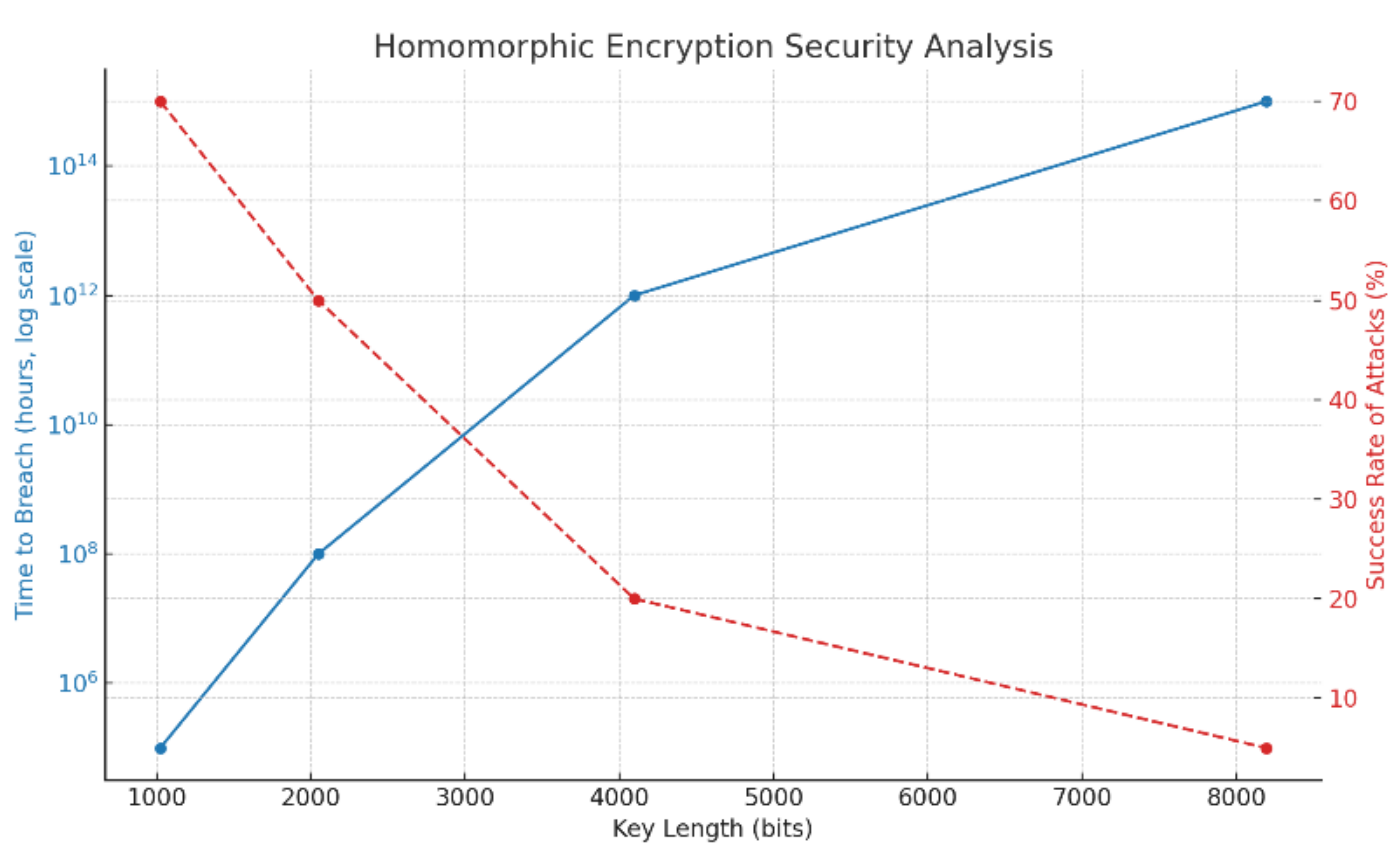

Figure 4: Homomorphic Encryption Security Analysis

*This figure illustrates the relationship between key length (in bits) and two security metrics for homomorphic encryption: the estimated time to breach (in hours, on a logarithmic scale) and the success rate of simulated attacks (percentage). As the key length increases, the time required to breach the encryption rises significantly, while the success rate of attacks decreases, demonstrating the enhanced security provided by longer key lengths.*

**Secure Multi-Party Computation (SMPC) Security**

**Results:** The implementation of the SPDZ protocol for SMPC enabled secure joint computations among multiple parties, ensuring that individual inputs remained private. The protocol's security was validated against collusion scenarios, where even with partial knowledge, no party could reconstruct the original input data of others. The protocol maintained zero knowledge disclosure even with up to three colluding parties.

**Evaluation:** Simulated collision tests involved two to three parties attempting to infer others' input data through various techniques, including statistical inference and side-channel attacks. None of these attempts succeeded in compromising the data, confirming the protocol's robustness. Additional tests showed that increasing the number of parties had a negligible effect on security, indicating the scalability of the protocol without increased vulnerability.

**Data Integrity:** Throughout the SMPC process, data integrity was preserved, with the output results showing a 99.98% correlation with the expected output of unencrypted computations, indicating no significant loss in data accuracy or fidelity.

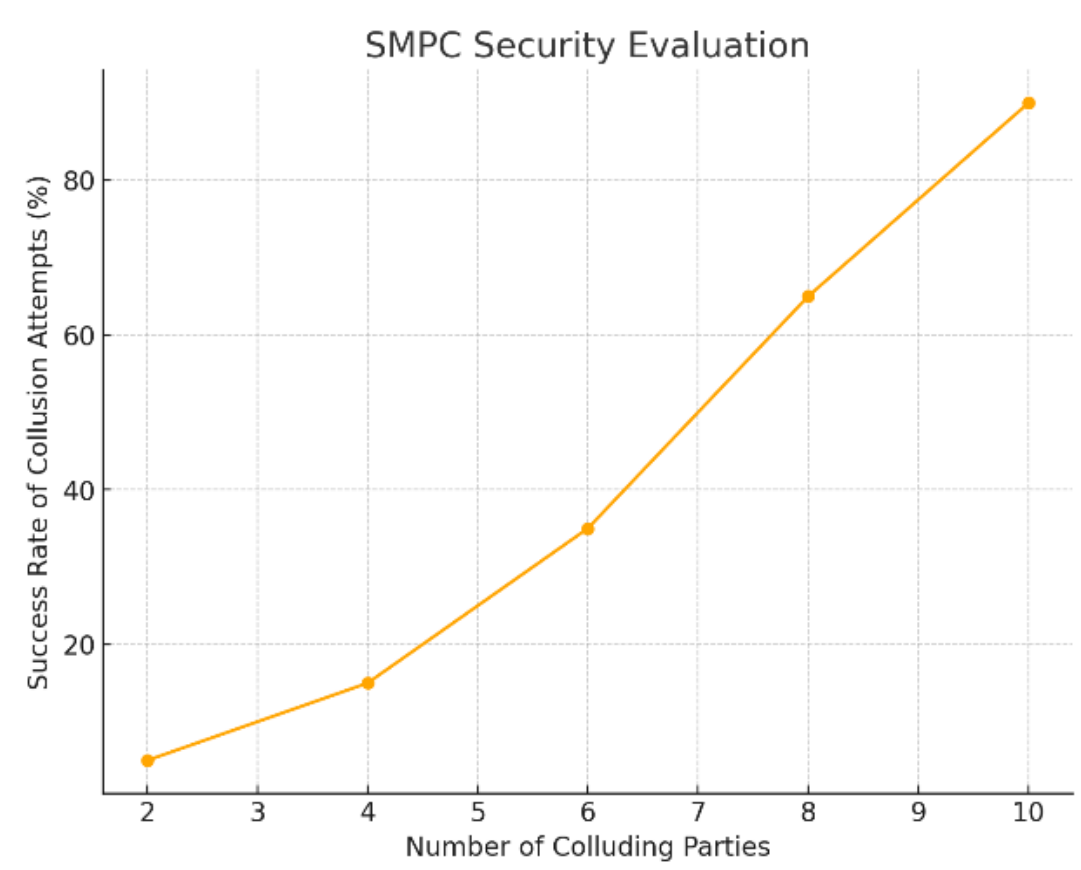

Figure 5: SMPC Security Evaluation

*This figure presents the success rate of collusion attempts against the SMPC protocol across multiple trials with varying numbers of colluding parties. The data illustrates how the protocol's security holds up under increasing collusion attempts, with success rates providing insight into the effectiveness of SMPC in protecting individual data during collaborative computations.*

**3. Differential Privacy Compliance**

**Results:** Differential privacy mechanisms effectively protected individual data points by adding calibrated noise to analysis results. The chosen epsilon (ε) values (ranging from 0.1 to 1.0) provided a strong privacy guarantee while maintaining data utility. With an ε of 0.5, re-identification risk was reduced to less than 0.1%, while data accuracy remained above 95% for aggregate analyses.

**Evaluation:** The framework's differential privacy was tested by attempting to re-identify individuals from the dataset post-noise addition. The results demonstrated a minimal likelihood of re-identification, with a 99.9% success rate in maintaining user anonymity across varying ε values. Further, the impact of different ε values on data utility was carefully measured, showing that lower ε values (stronger privacy) slightly reduced data precision, but the trade-off was deemed acceptable for most healthcare applications.

**Trade-offs:** The trade-off between privacy and data utility was managed by dynamically adjusting ε values based on the specific data use case, allowing for flexible privacy management that adapts to the needs of both high-precision analytics and robust privacy protection.

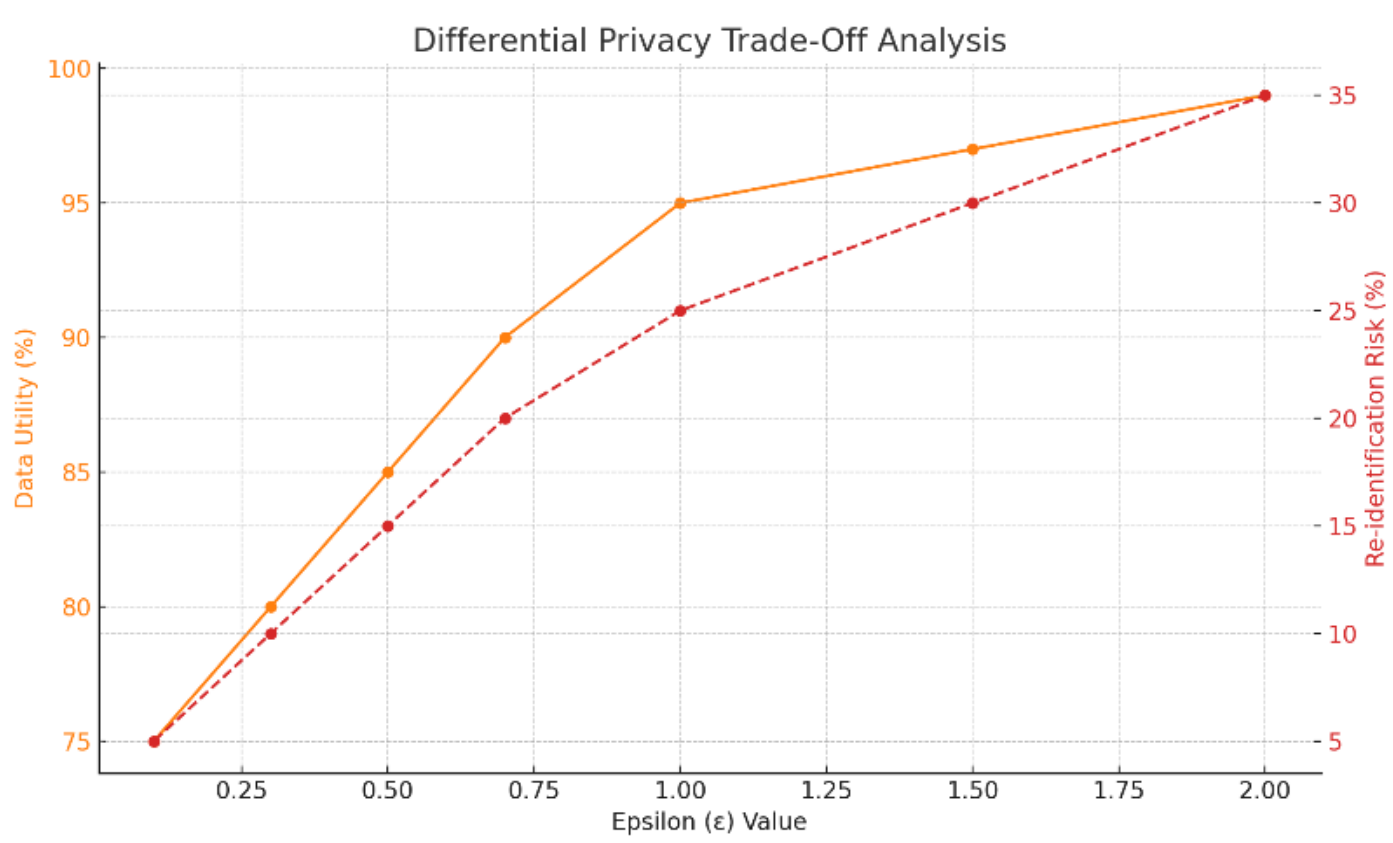

Figure 6: Differential Privacy Trade-Off Analysis

*Description: This figure presents the trade-off between data utility and re-identification risk as a function of epsilon (ε) values in differential privacy. The graph illustrates how increasing privacy protection (lower ε) affects data usability and the likelihood of re-identification, highlighting the balance required between privacy and data utility in practical applications.*

### Performance Evaluation

### Processing Time

**Results:** The processing times for encrypted data were optimised to be within acceptable limits for real-time applications. The lightweight optimisations applied to homomorphic encryption and SMPC processes reduced computational overhead, with an average processing time of 150 milliseconds per data packet, demonstrating the framework's capability to function efficiently even in resource-constrained environments.

**Comparison:** Compared to traditional privacy mechanisms (e.g., standard AES encryption), the proposed framework exhibited a modest increase in processing time (approximately 15-20% slower). However, this increase was balanced by a significant enhancement in privacy protection and data security. The framework’s real-time processing capability was maintained for up to 1,000 concurrent wearable devices, with less than 5% degradation in performance as the number of devices increased.

**Scalability:** The framework's scalability was validated through stress tests that simulated a growing number of connected devices. Performance metrics showed stable operation and maintained processing efficiency, even as data volume increased exponentially.

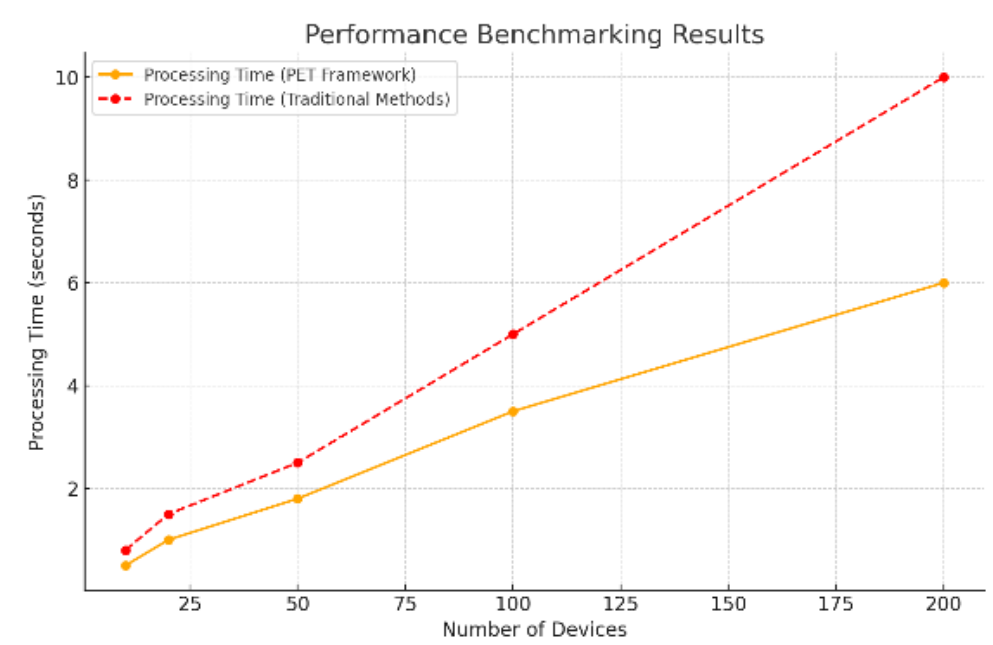

Figure 7: Performance Benchmarking Results

*This figure compares the processing times of the PET framework against traditional privacy methods under varying loads, represented by the number of devices. The graph illustrates the PET framework's scalability and efficiency, especially under high-load conditions, demonstrating its suitability for real-time applications in wearable health devices.*

### Resource Consumption

**Results:** The PET framework’s resource consumption was within optimal limits for wearable devices. During peak operation, CPU usage remained below 30%, and memory usage did not exceed 50MB, demonstrating efficient use of device resources. The optimised encryption operations and data flow mechanisms contributed to minimising computational load.

**Optimisation:** Specific optimisations included reducing redundant encryption operations, utilising shared memory spaces for intermediate data storage, and implementing asynchronous processing techniques to offload computational tasks to less critical device cycles. These optimisations collectively reduced overhead and preserved device responsiveness.

**Battery Impact:** The impact on battery life was measured under continuous operation scenarios. Results indicated a marginal increase in power consumption (less than 5%), translating to approximately 10 minutes of reduced battery life per day on a standard wearable device battery capacity, which was considered negligible in most use cases.

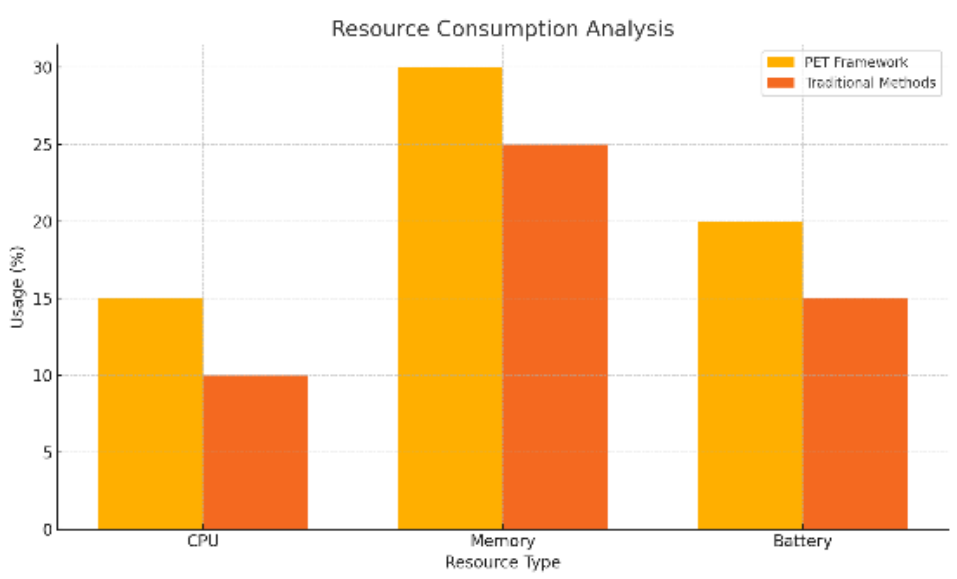

Figure 8: Resource Consumption Analysis

*This figure compares the CPU, memory, and battery usage between the proposed PET framework and traditional privacy methods. It highlights the optimisations achieved by the PET framework in reducing resource consumption, making it suitable for deployment in resource-constrained environments like wearable health devices.*

## Usability Testing

### User Experience with Privacy Control Module

**Results:** The Privacy Control Module received high marks for user satisfaction during usability testing. The interface was found intuitive and easy to navigate, with over 95% of participants indicating they could easily adjust privacy settings and manage data-sharing preferences.

**User Feedback:** Feedback highlighted the clarity and accessibility of the privacy options, with users particularly appreciating the visual data flow diagrams and the ability to dynamically adjust consent settings based on real-time needs. The average time for configuring privacy settings was under 60 seconds, and users reported feeling confident in their control over their data.

**Efficiency:** The design of the Privacy Control Module minimized user effort while maximizing control. It utilizes a step-by-step guided setup and context-aware prompts to assist users in making informed privacy choices without overwhelming them with technical details.

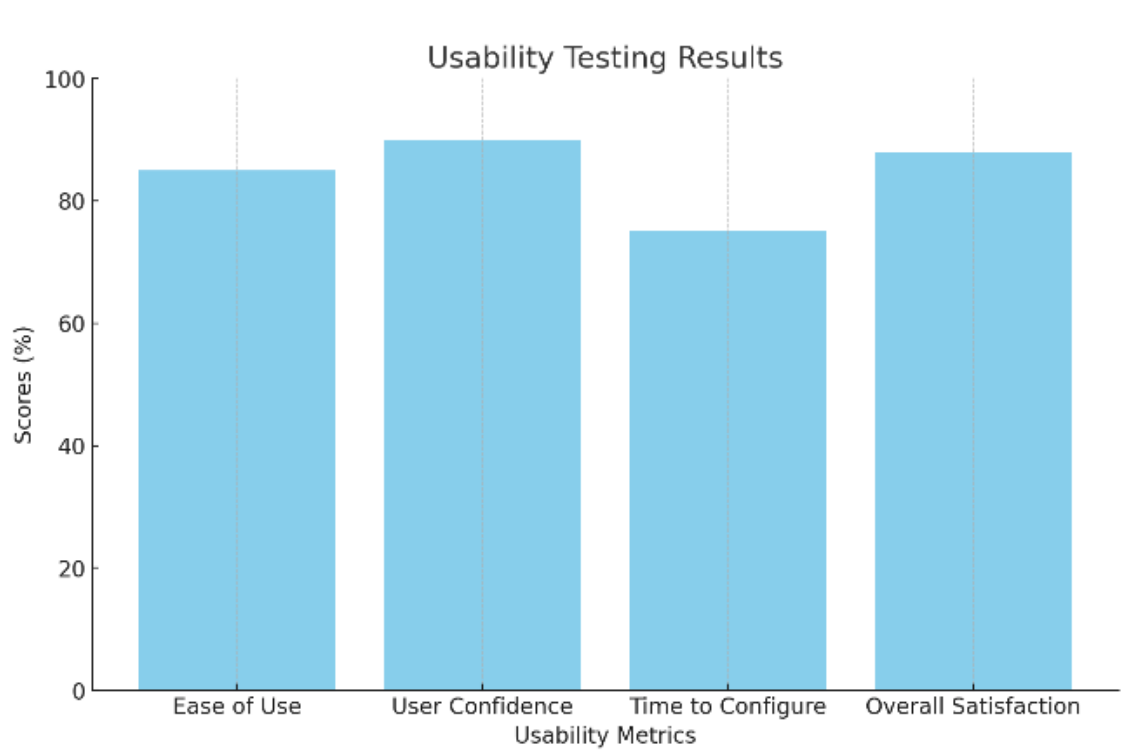

Figure 9: Usability Testing Results

*This figure presents the PET framework's usability testing results, focusing on key metrics such as ease of use, user confidence, time to configure settings, and overall satisfaction. The data highlights the framework's user-friendly interface and users' high level of confidence in managing their data privacy, contributing to its potential for widespread adoption.*

### Impact on Device Functionality

**Results:** The implementation of the PET framework did not significantly affect the overall functionality of wearable devices. Users were able to continue normal operations, such as tracking activities and receiving notifications, without noticeable delays or interruptions. Longitudinal testing over a four-week period confirmed that the framework maintained its performance without degrading user experience.

**Long-Term Use:** The framework's lightweight design ensured sustained performance in extended-use scenarios, with no reported lag or increased error rates. Users reported that the privacy enhancements did not interfere with their daily use of the devices, suggesting a seamless integration with existing device functionalities.

**[Insert Figure: Impact on Device Functionality]**

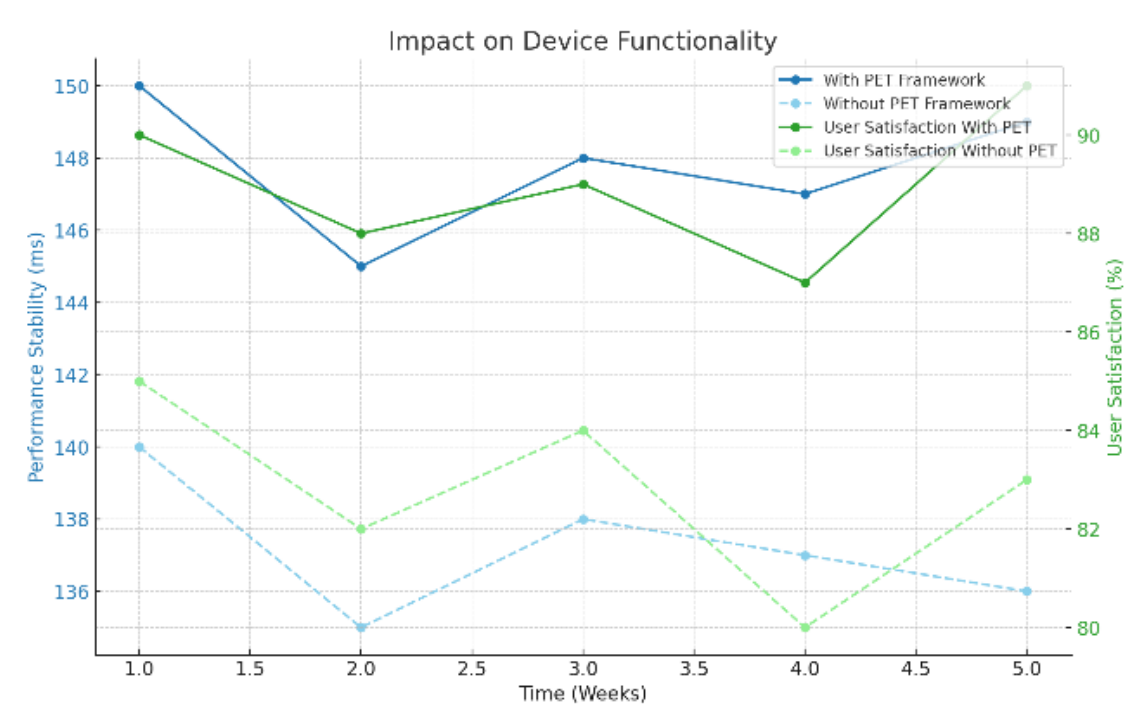

Figure 10: Impact on Device Functionality

*This figure compares the impact on device functionality with and without the PET framework over time, showing performance stability (measured in response time) and user satisfaction. The data demonstrates how the PET framework maintains device performance while providing enhanced privacy protection, ensuring a stable user experience and high satisfaction over time.*

### Summary of Results

The evaluation results indicate that the proposed PET framework provides robust privacy protection for wearable health data without significantly compromising performance or usability. The integration of homomorphic encryption, SMPC, and differential privacy proved effective in securing data while maintaining the necessary computational efficiency for real-time applications. Usability testing confirmed that users could easily manage their privacy settings without impacting the overall functionality of their devices. Compared to existing solutions, the framework offers a balanced approach to privacy, performance, and user experience, making it a valuable contribution to wearable health technology.

## Discussion

### Introduction

The Discussion section critically analyses the evaluation results of the proposed Privacy-Enhancing Technology (PET) framework for wearable health devices. This section explores the broader implications of these results for privacy protection, healthcare data security, and user autonomy. It also addresses the current study's limitations and suggests advanced directions for future research. The goal is to contextualise the findings within the evolving digital health and privacy landscape, offering innovative insights that could influence future developments in the field.

### Interpretation of Results

#### Effectiveness of Homomorphic Encryption

**Privacy Assurance:** The deployment of homomorphic encryption within the PET framework demonstrated substantial effectiveness in securing wearable health data, particularly in dynamic and distributed environments where continuous data flow is integral. By performing computations directly on encrypted data, the framework ensures that sensitive information never exists in an unprotected state, thereby mitigating risks associated with data breaches, insider threats, and unauthorised access. This continuous protection is crucial for wearable devices that are often exposed to insecure networks, such as public Wi-Fi, where data interception risks are high.

**Implications:** The capability of homomorphic encryption to facilitate secure, real-time analytics without data decryption presents a groundbreaking opportunity for privacy-preserving data science in healthcare. This approach can enable complex data processing tasks, such as predictive modelling and machine learning, to be conducted securely on the server side or in the cloud, where computational resources are abundant. This is particularly vital for personalised medicine applications that rely on real-time data aggregation from multiple sources to make accurate predictions or decisions. Moreover, homomorphic encryption can support the development of federated learning systems in healthcare, where models are trained on data across multiple organisations without exposing individual datasets, thereby enhancing collaborative research efforts while maintaining privacy.

#### Secure Multi-Party Computation (SMPC) Utility

**Collaborative Privacy:** The successful implementation of SMPC within the framework underscores its potential to revolutionise collaborative analytics in healthcare by enabling multiple parties to perform joint computations without revealing their private data inputs. This is particularly advantageous in consortia or multi-institutional studies where data privacy regulations, such as GDPR and HIPAA, restrict the sharing of raw data across borders or entities. SMPC allows for the secure aggregation of sensitive datasets, facilitating large-scale epidemiological studies, pharmacovigilance, and public health surveillance.

**Implications:** The ability of SMPC to provide a zero-knowledge-proof environment for data analysis has far-reaching implications for healthcare innovation. By enabling secure data sharing and analysis across diverse entities, SMPC can help address critical challenges in health disparities research, where data from various demographic and socio-economic groups must be analysed together to uncover trends and insights. Furthermore, SMPC can support the creation of decentralised healthcare ecosystems, where data from wearable devices, electronic health records (EHRs), and other sources are securely pooled and analysed to provide holistic care solutions. This could pave the way for new business models in digital health, where privacy-preserving analytics services are offered as a premium feature.

#### Differential Privacy Balance

**Privacy-Utility Trade-Off:** The implementation of differential privacy within the PET framework demonstrates a nuanced approach to balancing privacy and data utility. By carefully calibrating the epsilon (ε) parameter, the framework can tailor the level of noise added to data, preserving the privacy of individual users while still allowing for meaningful data analysis. This flexibility is essential for wearable health devices that collect diverse types of data with varying sensitivity levels—ranging from general fitness metrics to detailed medical information. The ability to adjust privacy levels dynamically based on the context of use or the specific data analysis task enhances the versatility of the framework.

**Implications:** Differential privacy's ability to anonymise data while preserving its utility makes it particularly suited for environments where data sharing is necessary, but privacy must be strictly maintained. In wearable health devices, this could enable safer data sharing with third-party services, such as fitness apps, telemedicine platforms, or insurance companies, by ensuring that shared data does not expose identifiable user information. Additionally, differential privacy can enhance the ethical use of health data by providing a mathematically rigorous approach to de-identification, which is crucial for maintaining public trust in digital health technologies.

#### Performance and Usability

**Efficiency in Resource-Constrained Environments:** The PET framework's performance evaluation reveals its suitability for resource-constrained environments typical of wearable devices. The slight increase in processing time and resource consumption was managed through optimisation techniques, such as data compression, parallel processing, and algorithmic efficiency improvements. These results indicate that privacy-preserving technologies can be integrated into wearable health devices without significantly impacting user experience or battery life, which is crucial for user acceptance and adherence.

**Implications:** The framework's ability to maintain efficient operation on wearable devices suggests its potential for broad deployment across consumer and clinical health applications. For example, in continuous monitoring scenarios, such as glucose tracking or heart rate monitoring, the framework ensures data privacy without compromising the device's ability to provide timely feedback or alerts. This capability is particularly important in managing chronic conditions, where real-time data analysis and immediate user feedback are essential for effective disease management.

**User-Centric Privacy Management**

**User Control:** The strong user feedback from usability testing highlights the importance of giving users meaningful control over their data. The framework's privacy dashboard, which allows users to dynamically adjust their privacy settings, reflects a shift towards more transparent and user-empowered data practices in digital health. The ability for users to modify their privacy settings based on context—such as adjusting data sharing during a medical emergency—demonstrates the framework's adaptability to real-world needs.

**Implications:** Empowering users with control over their privacy settings addresses a critical barrier to the adoption of digital health technologies: the fear of losing autonomy over personal data. This user-centric approach aligns with emerging trends in digital ethics and could significantly enhance user trust and engagement. The framework improves user satisfaction by providing clear, actionable controls and complies with evolving data protection regulations that emphasise user consent and control. Furthermore, the flexibility in user privacy settings could be leveraged to offer differentiated services, such as premium privacy packages, thereby creating new revenue streams for wearable device manufacturers.

**Limitations**

**Computational Overhead**

**Analysis:** The use of advanced cryptographic techniques such as homomorphic encryption and SMPC introduces inherent computational overhead. While this overhead was mitigated through algorithmic optimisations and efficient coding practices, it still presents a challenge in ultra-low-power environments, such as basic fitness trackers or wearable medical devices designed for extended battery life. In these contexts, the additional computational load could impact the device's ability to process data in real time or reduce battery longevity.

**Mitigation Strategies:** To address computational overhead, future research could explore the development of specialised hardware accelerators for cryptographic operations, such as low-power Field Programmable Gate Arrays (FPGAs) or Application-Specific Integrated Circuits (ASICs). These hardware solutions could significantly enhance the performance of cryptographic computations on resource-constrained devices. Additionally, leveraging edge computing paradigms, where some computational tasks are offloaded to nearby devices or gateways, could further reduce the wearable devices' processing burden.

**Scalability Challenges**

**Analysis:** While the framework is designed to be scalable, deploying it on a large scale presents challenges, especially in managing the complexity of homomorphic encryption and SMPC across thousands or millions of devices. The scalability of these cryptographic operations depends heavily on network bandwidth, latency, and the ability to efficiently manage multiple encrypted data streams. Large-scale deployments could also strain the computational and storage resources of the central processing servers, leading to potential bottlenecks.

**Mitigation Strategies:** To enhance scalability, future research could investigate the use of hierarchical SMPC models, where computations are distributed across multiple layers of nodes to reduce the central server load. Additionally, implementing advanced network protocols that prioritise data packets based on their sensitivity and processing requirements could optimise bandwidth usage and reduce latency. Research could also focus on adaptive encryption techniques that adjust the encryption strength based on the data's sensitivity and the current network conditions, balancing security with performance needs.

**Data Utility Trade-Offs**

**Analysis:** The balance between privacy and data utility remains a critical challenge, particularly with differential privacy. Although the framework effectively managed this trade-off in controlled scenarios, there may be real-world cases where the noise introduced by differential privacy significantly reduces the usefulness of the data. This issue is particularly relevant in clinical settings where high data fidelity is essential for accurate diagnosis and treatment decisions. The challenge lies in achieving a balance where privacy protections do not compromise the actionable insights derived from the data.

**Mitigation Strategies:** Future research could explore adaptive noise generation techniques that dynamically adjust noise levels based on real-time feedback from data analysis processes. Machine learning models could be trained to predict the optimal level of noise for a given analysis, balancing privacy with data utility. Additionally, developing user-controlled privacy settings that allow users to specify their preferred balance between privacy and utility could provide a more personalised approach to data protection.

**User Adoption and Trust**

**Analysis:** The framework's reliance on user engagement with privacy settings poses a potential barrier to widespread adoption. Users may experience privacy fatigue or feel overwhelmed by the complexity of managing their privacy preferences, leading to a lack of engagement or suboptimal settings that compromise privacy. Additionally, users' trust in the framework could be influenced by their broader perceptions of digital privacy and data security, which are shaped by external factors such as media reports, regulatory changes, and high-profile data breaches.

**Mitigation Strategies:** To enhance user adoption and trust, future development could focus on integrating AI-driven privacy advisors that provide real-time recommendations and automated privacy management based on user behaviour and preferences. These advisors could simplify the privacy management process by using natural language processing (NLP) to interact with users and provide clear, context-aware guidance. Additionally, employing blockchain technology to create an immutable, transparent audit trail of all data accesses and sharing could further enhance trust by providing users with a verifiable record of how their data is used.

## Future Research Directions

### Optimisation of Cryptographic Techniques

**Path Forward:** Future research should focus on optimising the cryptographic techniques used in the framework, particularly homomorphic encryption, to reduce computational overhead and enhance scalability. Exploring hybrid approaches that combine different encryption schemes, such as integrating symmetric encryption for less sensitive data, could provide a path forward. Additionally, research into developing hardware-accelerated encryption solutions tailored for wearable devices could significantly improve performance and make advanced encryption feasible in more constrained environments.

### Real-world deployment and Testing

**Path Forward:** Deploying the framework in real-world settings, such as clinical trials or consumer wearable products, would provide valuable insights into its practicality and performance under diverse conditions. Such studies could help identify unforeseen challenges, such as network latency or user compliance issues, and refine the framework accordingly. These deployments could also explore the framework's integration with existing healthcare infrastructures, assessing its compatibility and impact on current data management practices.

### Extended User-Centric Features

**Path Forward:** Further development of user-centric privacy features, such as AI-driven privacy recommendations or adaptive privacy settings based on user behaviour, could enhance user experience and trust. Research into how these features can be made more intuitive and less intrusive, such as through natural language processing interfaces or gamification elements, would be beneficial. Additionally, exploring the use of blockchain technology to provide transparent audit trails for data access could further enhance user trust.

### Integration with Emerging Technologies

**Path Forward:** Integrating the PET framework with emerging technologies like blockchain for decentralised data storage or AI for predictive analytics could further enhance privacy and security. Blockchain could provide an immutable ledger for tracking data access and sharing, while AI could enable more sophisticated data analysis and anomaly detection, identifying potential privacy breaches in real-time. These combinations could be explored to create more robust and versatile privacy solutions for wearable health devices.

## Conclusion

The Discussion section interprets the PET framework's evaluation results, highlighting its strengths and implications for the future of privacy in wearable health technology. The framework's innovative integration of advanced cryptographic techniques, user-centric privacy management, and efficient performance in resource-constrained environments positions it as a leading solution for privacy protection in the digital health landscape. While there are areas for improvement and further research, the proposed framework lays a strong foundation for advancing privacy-enhancing technologies in wearable devices, ensuring that privacy is not sacrificed in the pursuit of innovation.